\documentclass[aps,prb,preprint,psfig,groupedaddress]{revtex4}
\usepackage{graphicx}
\usepackage{bm}

\newcommand{\mum}{$\mu\rm{m}$}
\newcommand{\tc}{$T_{\rm{C}}$}
\newcommand{\BS}{Bi$_2$Se$_3$}
\newcommand{\vF}{$v_{\rm{F}}$}

\newcommand{\lel}{$l_{\rm{el}}$}

\newcommand{\Lphi}{$L_{{\phi}}$}
\newcommand{\xiN}{$\xi_{\rm{N}}$} 
\newcommand{\Fs}{$\tilde{F}_{\sigma}$}

\renewcommand{\vec}[1]{\mbox{\boldmath$1$}}

\newcounter{lastnote}

\def\bc{\begin{center}}
\def\ec{\end{center}}
\def\be{\begin{equation}}
\def\ee{\end{equation}}
\renewcommand{\vec}[1]{\mbox{\boldmath$1$}}

\usepackage{amsmath,amssymb}

\begin{document}
\title{Observation of the superconducting proximity effect and possible evidence for Pearl vortices in a candidate topological insulator}
\author{Duming Zhang, Jian Wang, Ashley M. DaSilva, Joon Sue Lee, Humberto R. Gutierrez, Moses H. W. Chan, Jainendra Jain, Nitin Samarth}
\email{nsamarth@psu.edu}
\affiliation{The Center for Nanoscale Science and Department of Physics, The Pennsylvania State University, University Park, Pennsylvania 16802-6300, USA}

\date{\today}
\begin{abstract}
We report the observation of the superconducting proximity effect in nanoribbons of a candidate topological insulator (\BS) which is interfaced with superconducting (tungsten) contacts. We observe a supercurrent and multiple Andreev reflections for channel lengths that are much longer than the inelastic and diffusive thermal lengths deduced from normal state transport. This suggests that the proximity effect couples preferentially to a ballistic surface transport channel, even in the presence of a coexisting diffusive bulk channel. When a magnetic field is applied perpendicular to the plane of the nanoribbon, we observe magnetoresistance oscillations that are periodic in magnetic field. Quantitative comparison with a model of vortex blockade relates the occurrence of these oscillations to the formation of Pearl vortices in the region of proximity induced superconductivity. 
\end{abstract}
\maketitle

\section{introduction}
Proximity-induced superconductivity in superconductor-normal conductor (SN) junctions has a long history of theoretical and experimental study\cite{Pannetier00}. Recently, the superconducting proximity effect has attracted renewed theoretical attention within the context of topological insulators (TIs), materials wherein topologically protected, spin-polarized surface states are created by the combination of strong spin-orbit coupling and time reversal symmetry \cite{Hasan2010,Qi2010}. Interfacing a TI with a conventional superconductor is of fundamental interest for a variety of reasons. At such interfaces, theory predicts the formation of zero-energy mode quasiparticles that are condensed matter analogs of elementary fermionic excitations envisioned by Majorana but not yet observed in Nature \cite{Fu2008,Sau2010,Franz2011}. Furthermore, the ``locking'' of spin and momentum in TI surface states raises important theoretical questions about the nature of the proximity effect at TI/superconductor junctions \cite{Stanescu2010} and recent experiments suggest that the measurement geometry could have a significant influence on the observed phenomena \cite{Wang_2011B}. Evidence for the proximity effect was provided in an early study \cite{Kasumov} that interfaced a superconductor with Bi$_{1-x}$Sb$_x$, a material now recognized \cite{Fu2007,Hsieh2008} as a 3D TI. More recently \cite{Sacepe2011}, a supercurrent was observed in thin exfoliated samples of another TI (\BS). Bulk superconductivity has also been seen in compounds derived from a parent TI  \cite{hor10a,wray10}. However, systematic studies of proximity-induced superconductivity in candidate TIs are still in their infancy.

In this paper, we discuss measurements of the proximity effect in mesoscopic \BS~channels interfaced with superconducting tungsten (W) leads. Our principal aim is to report the observation of two new experimental results. First, we show that even in non-ideal TI samples with bulk conduction, ballistic transport in the surface states can manifest through the persistence of a supercurrent and multiple Andreev reflections over significantly longer distances than the phase breaking and diffusive thermal lengths deduced from (bulk dominated) normal state transport. Second, at temperatures above the onset of complete superconductivity, we observe magnetoresistance (MR) oscillations that cannot be understood using conventional scenarios such as fluxoid quantization or the Aharonov-Bohm effect. Instead, the data are {\it quantitatively} explained using a recent model \cite{Pekker} that relates MR oscillations in superconducting channels to the ``Weber blockade'' of Pearl vortices. The combined observation of proximity induced superconductivity and vortices in a TI-superconductor configuration could be relevant in the ongoing search for Majorana fermions \cite{Fu2008,Sau2010,Franz2011}.

\section{Synthesis and characterization of \BS~nanoribbons}
The \BS~nanoribbons studied here were synthesized via gold catalyzed vapor-liquid-solid mechanism using a horizontal
tube furnace\cite{Peng10,Kong10}. The source material, \BS~shots (Alfa Aesar, 99.999\%%
), was placed in the center of the hot furnace where the temperature was kept at $\sim 530\,^{\circ}$C during the growth. A Si (100) substrate dispersed with gold catalyst particles ($\sim 20$ nm in diameter) was placed downstream at the cold region zone. The tube was flushed with Ar several times before the growth to remove residual oxygen. A 60 sccm Ar flow was kept at 1 Torr at the base pressure of $\sim 10$ mTorr during the growth. The furnace was cooled down to room temperature after a 1 hour and 30 minute growth. Figure 1a shows a scanning electron microscopy (SEM) image of an as-grown sample. A typical growth usually produces nanoribbons with thickness $60 \lesssim t \lesssim 100$ nm, width $200 \lesssim w \lesssim 500$ nm and length $2 \lesssim L \lesssim 30$ \mum.  Figure 1b shows a high resolution transmission electron microscopy (TEM) image from the edge of a typical nanoribbon. The interplanar distance along the nanoribbon growth direction is 0.21 nm, which is consistent with the \BS~lattice constant ($a = b = 0.4140$ nm, $c = 2.8636$ nm). Figure 1c is a selected area electron diffraction pattern from the same ribbon: all the nanoribbons we have studied so far show a growth direction along $[11 \bar{2} 0]$. The clear lattice fringes and the electron diffraction pattern indicate that these ribbons are single crystals with little disorder.

We obtained additional confirmation about the crystalline phase of the samples using Raman spectroscopy of individual \BS~nanoribbons. Figure 1d shows a typical room temperature Raman spectrum from a single \BS~nanoribbon supported over one of the holes in a TEM grid; the data were taken with 514.5 nm excitation and the geometry ensures that the back scattered light only originates from the nanoribbon of interest. The incident radiation was polarized parallel to the nanoribbon growth direction and the scattered radiation was unpolarized. The measurements were performed with a very low incident laser power ($\sim 10\,\mu$W with a spot size of $\sim 1\,\mu$m) to avoid sample overheating. Three characteristic Raman modes are observed: $\sim$ 72 cm$^{-1}$, out of plane vibration mode;  $\sim$132 cm$^{-1}$, in plane vibration mode; and $\sim$ 173 cm$^{-1}$, out of plane vibration mode. All these modes are consistent with those observed from bulk \BS \cite{Richter77}.

We fabricated electrical transport devices by transferring as-grown \BS~nanoribbons to a Si substrate with a 1 \mum~ thick Si$_3$N$_4$ insulating layer and then depositing superconducting tungsten and/or normal platinum (Pt) electrodes using a dual beam focused ion beam (FIB) etching and deposition system. The tungsten contacts are contaminated with carbon and gallium, and consequently have a high superconducting transition temperature (4 K$ \lesssim$ \tc  $\lesssim$ 5 K) and a large critical field ($H_c \gtrsim 70$ kOe)  \cite{Wang_2010}. The correspondingly large superconducting energy gap ($\Delta \sim 0.7$ meV) allows good energy resolution in differential conductance measurements. Furthermore, the large critical field lends such devices to explorations of possible vortex formation in regions of proximity-induced superconductivity. The FIB method also has the advantage of forming good ohmic contacts with \BS~nanoribbons, albeit at the price of localized damage in the contact area. TEM measurements of test structures (Fig. 2a) show that the nanoribbons remain in the single crystal \BS~phase after FIB deposition, apart from the region immediately underneath the tungsten contacts. The typical electrode is about $200-400$ nm wide and $\sim 50$ nm high. X-ray energy dispersive profile scans in scanning TEM studies (Fig. 2b) shows that the tungsten spreads about 250 nm from the visible edge of the contact. We conservatively estimate that the contacts have a resistance-area product $RA \sim 10^{-10}~\Omega$m$^2$. To rule out the possible shorting between closely spaced tungsten contacts, we also carried out a control measurement by depositing two tungsten strips with a visible edge-to-edge separation of 460 nm on a Si$_3$N$_4$ substrate; we found that the (two-probe) resistance of this configuration was larger than 0.5 M$\Omega$ at $T = 500$ mK, well below the superconducting transition temperature of the tungsten contacts. To further rule out insidious spreading effects that could occur in a nanostructure geometry, we also examined (undoped) ZnSe nanowires with tungsten contacts and again found no signs of superconducting shorts below the transition temperature of the contacts.

We measured a total of 6 \BS~nanoribbon devices with tungsten contacts; their characteristics are summarized in Table 1. In addition, we also measured two devices with non-superconducting (Pt) contacts. Although the nanoribbons were all fabricated in the same synthesis run, their dimensions and electrical properties (resistivity) show an inhomogeneous distribution.We are unable to determine the carrier density and carrier mobility because the nanoribbons are too narrow to carry out Hall effect measurements; however, the electrical resistivity of all the devices ($\sim$ 1 m$\Omega\cdot$cm) is of the same order as typical values observed in thin films of \BS, suggesting similar carrier densities ($n \sim 10^{19}~{\rm{cm}}^{-3}$). We measured the bias-dependent differential conductance in single \BS~nanoribbon devices using a dc current source and a lock-in amplifier at a frequency of 97 Hz over a temperature range 0.5 K $\leq  T \leq$ 6 K and in magnetic fields up to 80 kOe. The measurements were performed in both four probe and two probe geometries; the latter measurements are carried out in a ``pseudo-four-probe" scheme, with two wires on each contact pad. 

\section{Transport measurements with normal metal (Pt) contacts}
Before we discuss transport measurements of \BS~ nanoribbons with superconducting electrodes, we first study transport properties of two devices with non-superconducting electrodes in the four probe geometry. Figures 3a and 3b plot the temperature dependent conductivity for device G and H respectively, showing a $\ln (T)$ variation, consistent with 2D quantum corrections due to weak anti-localization and electron-electron interactions\cite{Wang_2011}. Detailed fitting of $\sigma$ vs $T$ becomes more complicated because of the many fitting parameters: as we show below, fitting the magnetoconductivity to the weak localization theory suggests that there are multiple channels participating in the quantum corrections. Since the slope of $\sigma$ vs. $\ln (T)$ is strongly affected by the dependence of $L_{\phi}(T)$, it is not appropriate to assume all channels contribute equally, as this assumes the coherence length for each of the channels is the same as the others. Figures 3c and 3d show the magnetoconductivity for device G and H at $T = 3.0$ K (squares) and $T = 2.0$ K (circles) in fields perpendicular to the nanoribbon plane. We fit the magnetoconductivity using the Hikami-Larkin-Nagaoka\cite{HLN_1980} equation:
\begin{equation}
\Delta\sigma_{\rm{WL}}(H)-\Delta\sigma_{\rm{WL}}(0) = \alpha \frac{e^2}{2\pi ^2\hbar}\left ( \Psi \left( \frac{1}{2}+\frac{\hbar c}{4eL_{\phi}^2 H}\right) - \ln \left( \frac{\hbar c}{4eL_{\phi}^2 H}\right) \right) \rm{,}
\end{equation}
\newline
where $\alpha$ depends on the dominant scattering mechanism and $\Psi(x)$ is the digamma function.  $\alpha = 1$ in the limit of weak spin orbit and magnetic scattering; $\alpha = -1/2$ in the limit of strong spin orbit scattering and weak magnetic scattering; and $\alpha = 0$ when magnetic scattering is strong. For completeness, we also include the 2D correction from electron-electron interactions\cite{Lee_1985} given by:
\begin{equation}
\Delta\sigma_{\rm{EEI}}(T,H)-\Delta\sigma_{\rm{EEI}}(T,0) = -\frac{e^2}{4\pi ^2\hbar}\tilde{F}_{\sigma}g_2 (T,H)\rm{,}
\end{equation}
\newline
where
\begin{equation}
g_2 (T,H) = \int_0^\infty \mathrm{d}\Omega \ln\left\lvert 1-\left( \frac{g\mu_B H/k_B T}{\Omega}\right) ^2 \right\rvert \frac{d^2}{d\Omega^2} \frac{\Omega}{e^{\Omega}-1}\rm{,}
\end{equation}
and \Fs~is a function of the average of the static screened Coulomb interaction over the Fermi surface.  The solid lines in Figs. 3c and 3d  are fits to the above equations, using as fitting parameters the phase breaking length \Lphi, the parameter $\alpha$ and the screening parameter \Fs. For device G, the fits yield \Lphi~$\sim 90.77$ nm, $\alpha = -1.30$ and \Fs~$= 0.0$ at $T = 3.0$ K and \Lphi~$\sim 137.2$ nm, $\alpha = -1.34$ and \Fs~$= 0.0$ at $T = 2.0$ K. For device H, the fits yield \Lphi~$\sim 35.5$ nm, $\alpha = -1.36$ and \Fs~$= 0.32$ at $T = 3.0$ K; \Lphi~$\sim 47.0$ nm, $\alpha = -1.29$ and \Fs~$= 0.36$ at $T = 2.0$ K. Note that the values of the phase breaking length are somewhat smaller than typically seen in thin film samples of \BS \cite{Wang_2011}. 

\section{Transport measurements with superconducting (tungsten) contacts: observation of a supercurrent and multiple Andreev reflections}

Figure 4a shows the temperature dependence of the two probe zero bias differential resistance ($dV/dI$) of device A with width $w = 600$ nm and thickness $t \sim 60$ nm. The resistance is measured between contacts with edge-to-edge separation $L = 1.08$ \mum (SEM image in the inset). Spreading of the tungsten contacts implies a conservatively estimated \BS~channel length of $\sim 580$ nm. The figure shows the onset of the proximity effect at $T \sim 4.7$ K when the W contacts become superconducting, eventually transitioning to a zero resistance supercurrent state at $T \leq 2$ K. Figure 4b plots the $I-V$ characteristics of this device at different temperatures, showing a critical current $I_c = 1.1\,\mu$A at $T = 500$ mK. The product $I_c R_N$ (where $R_N$ is the normal state resistance) has a small value ($\sim 165\,\mu$V $<< \frac{\Delta}{e}$), similar to observations in exfoliated \BS~layers \cite{Sacepe2011}. As expected, $I_c$ decreases with increasing temperature until the supercurrent is completely suppressed at $T \geq 2$ K. 

Figure 4c plots $dI/dV$ vs. $V$ at $T = 500$ mK, revealing ``subharmonic gap structure'' due to multiple Andreev reflections at $V = \frac{2\Delta}{ne}$, where $\Delta$ is the gap of the W electrodes and $n$ is an integer \cite{octavio1983,Cuevas2006,Xiang2006,Heersche}. The arrows identify a consistent series of subharmonic peaks in $dI/dV$ corresponding to $n = 2,4,8$ (Fig. 1d) for a superconducting energy gap $\Delta = 0.669$ meV. Using the BCS relation $\Delta = 1.76 k_B T_C$, this corresponds to \tc $= 4.41$ K, which is close to \tc $= 4.7$ K obtained from the temperature dependent zero bias differential resistance. The absence of some values of $n$ is not understood, but is not an uncommon occurrence \cite{Xiang2006}. The peak assignment in Fig. 4c results in an anomalous location of the prominent $n = 1$ peak at 1.194 meV instead of the expected location $V = \frac{\Delta}{e} = 1.338$ meV (indicated by the arrow in Fig. 4c). We attribute this to heating at higher bias, noting that similar anomalies have been seen in other SNS devices \cite{Xiang2006}. The origin of the peak identified as $\Delta ' = 0.932$ meV is unclear; it could be a signature of a quasiparticle minigap $\Delta_{\rm{g}}$ that is expected to arise when $L \sim \sqrt{\hbar D/\Delta} < L_\phi$ (i.e. when the Thouless energy roughly equals the superconducting gap of the S contacts) \cite{Cuevas2006}. The data in Figs. 1a-d provide a clear indication of proximity-induced superconductivity in \BS. In Figs. 4e and 4f, we map out the detailed variation of $dI/dV$ as a function of temperature and magnetic field, respectively. These plots show that -- as expected from the temperature and field dependence of the superconducting gap of the tungsten electrodes -- the multiple Andreev reflection features and $I_c$ are quenched with increasing temperature and field.

Observation of a supercurrent and multiple Andreev reflections in an SNS device requires a normal channel length $L$ shorter than two principal length scales: the thermal length \xiN~which characterizes the spatial decay of the pair amplitude in the N channel and the electron phase-breaking length \Lphi. At low temperatures, the cut-off is typically determined by \Lphi. For diffusive channels ($L >$ \lel), \xiN $= \sqrt{\hbar D/2 \pi k_B T}$, while for ballistic channels ($L < $ \lel), \xiN~$= \hbar v_F/2 \pi k_B T$. Here, $D$ is the diffusion constant, $v_F$ is the Fermi velocity, $k_B$ is the Boltzmann constant and $T$ is temperature. Since we do not directly know the carrier density $n$ in our samples, we estimate \lel~and \xiN~based upon some reasonable assumptions. In the normal state, the magnitudes of the resistivity and magnetoresistance of this nanoribbon ($\sim 0.51\,\rm{m}\Omega\cdot$cm) are similar to measurements in \BS~thin films\cite{Richardella} with a carrier density $n \sim 10^{19}\,\rm{cm}^3$. We thus use this carrier density as an upper limit for $n$ and use the Drude model (\lel~$ =v_{\rm{F}}m_{\rm{eff}}/\rho ne^2$) to yield \lel~$\sim$ 50 nm i.e. transport in the channel must be diffusive. Using the experimentally reported effective mass for \BS, $m_{\rm{eff}} \approx 0.18 m_e$, at roughly this carrier density and for transport normal to the c-axis \cite{Hyde_1974}, we estimate a Fermi velocity \vF~$\sim 4 \times 10^5~\rm{m} {\rm{s}}^{-1}$ from \vF~$=\frac{\hbar}{m_{\rm{eff}}}(3\pi^2 n)^{1/3}$ and a bulk diffusion constant $D \sim 7 \times 10^{-3}~{\rm{m}}^2 {\rm{s}}^{-1}$ from $D=v_{\rm{F}}l_{\rm{el}}/3$. This yields \xiN~$\sim 70$ nm at $T = 1.8$ K. We note that even if the assumed nanoribbon carrier concentration was an order of magnitude lower (i.e. $10^{18}~\rm{cm}^{-3}$), the corresponding mean free path \lel~would be $\sim 250$ nm and thermal length \xiN~$\sim100$ nm at 1.8 K, both of which are still smaller than the channel length. Next, we estimate a lower bound for \Lphi~by extracting the electron inelastic length from weak antilocalization fits to the magnetoconductance of \BS~nanoribbons with normal metal (Pt) contacts. (The large $H_c$ for the W contacts precludes a direct measurement of the low temperature magnetoconductance in samples with superconducting contacts.) In the two devices measured (Fig. 3c and 3d), we find \Lphi $\lesssim 140$ nm at $T = 2$ K. 

Thus, our analysis shows that we are observing a supercurrent and multiple Andreev reflections in SNS junctions with N channel lengths that are significantly longer than the values for \xiN~and \Lphi~deduced from the diffusive normal state transport. The simplest explanation for this apparent inconsistency is the coexistence of ballistic (surface) and diffusive (bulk) transport channels. We speculate that the latter may dominate the transport in the normal state, but that the proximity-induced superconductivity preferentially occurs in the ballistic surface channel. The validity of this conjecture can be checked by estimating \xiN~in the ballistic limit, using the value of the Fermi velocity \vF~$\sim 5 \times 10^5 \rm{m} {\rm{s}}^{-1}$ for the TI surface state in \BS \cite{Zhangh09}; this yields a value \xiN $_{\rm(clean)}= \frac{\hbar v_F}{2 \pi k_B T} \sim 340$ nm at $T = 1.8$ K, which is more consistent with the observation of a supercurrent at comparable channel lengths.

Our hypothesis of a ballistic surface channel that preferentially supports proximity-induced superconductivity is bolstered by transport measurements in three other \BS~nanoribbon devices. Figure 5 presents data taken on device B, a \BS~nanoribbon of thickness $t = 60$ nm and width $w = 430$ nm. The normal state resistivity of this nanoribbon (1.37 m$\Omega\cdot$cm) is almost three times greater than device A, yielding shorter estimated elastic and thermal lengths (\lel~$\sim$ 20 nm, \xiN~$\sim$ 40 nm). This device has two measurable channels (B1 and B2 in the inset to Fig. 5a) with conservatively estimated channel lengths $L_{B1} \sim 0.44\,\mu$m and $L_{B2} \sim 1.05\,\mu$m, respectively. Figures 5a-b show the onset of the proximity effect in these channels, while Figs. 5c-d show the bias-dependent differential conductance. The presence of multiple Andreev reflections in channel B1 and their absence in channel B2 are both consistent with our earlier estimate of $\xi_{\rm{N(clean)}} \sim 340$ nm for a clean surface channel since $L_{B1} \sim \xi_{\rm{N(clean)}} \ll L_{B2}$. We have also observed subharmonic gap structure in two additional nanoribbons (not shown). The overall behavior is similar to that discussed above.

\section{Transport measurements with superconducting (tungsten) contacts: anomalous magnetoresistance oscillations}
We now discuss the occurrence of unusual MR oscillations in \BS~nanoribbons under conditions where the contacts become superconducting but where a supercurrent has not yet been established. In order to make consistent comparisons, we carried out both sets of measurements in the two probe geometry on device B, recalling that the electrode configuration shown in Fig. 5a precludes a true four probe measurement for the shorter channel. As a check, we also measured channel B2 in the four
probe geometry and found identical behavior as the two probe measurements. Figure 6a shows the MR for channel B1 at $T$ = 2.2 K with a magnetic field applied perpendicular to the plane of the nanoribbon. At low field ($|H| < 0.6$ kOe), we observe a sharp cusp characteristic of weak anti-localization, similar to earlier results from \BS nanoribbons contacted with normal
electrodes. However, at higher fields, we find well defined and reproducible MR oscillations superimposed on a smooth background. The inset to Fig. 6a plots the magnetic field position of both peaks and valleys. If we exclude the low field region, we find that the MR oscillations are periodic in $H$ with a period $\Delta H \sim 3.22$ kOe. Figure 6b shows that the oscillation amplitude varies non-monotonically with temperature, reaching a maximum in the vicinity of $T$ = 2.2 K and vanishing at both lower and higher temperatures. Similar behavior is observed in channel B2 except with a smaller period $\Delta H \sim 1.44$ kOe (Figs. 6c and 6d). Similar oscillatory MR has also been observed in 2 other devices. For example, Fig. 7 shows the MR of device D in three primary field orientations: perpendicular to the plane of the nanoribbon (Fig. 7a and 7b), parallel to the nanoribbon axis (Fig. 7c), and parallel to the plane but perpendicular to the axis of the nanoribbon (Fig. 7d). While similar MR oscillations that are periodic ($\Delta H \sim 1.60$ kOe) at higher field were observed when the field is perpendicular to the nanoribbon plane, they were not observed in the other two primary field directions.

The observed MR oscillations cannot be explained using standard scenarios. The Shubnikov-de Haas effect is ruled out since the oscillations are not periodic in $1/H$. The Aharonov-Bohm effect of surface currents \cite{Peng10} is also ruled out for several reasons: the oscillations are only observed when the magnetic field is perpendicular to the nanoribbon plane and not when it is parallel; the oscillation amplitude has a non-monotonic temperature dependence; finally, MR oscillations are absent when using normal metal contacts (e.g. Pt down to $T = 2$ K or W contacts above the superconducting transition temperature), clearly indicating that important role of superconductivity. We note that the Aharonov-Bohm oscillations were not observed in device D in parallel field due to the dominance of the normal transport in a long channel ($L \gg \xi_{\rm{N(clean)}}$) even when the electrodes are in the superconducting state. We also discount another conventional explanation for the oscillatory MR, namely fluxoid quantization which requires a multiply connected geometry where flux can penetrate normal regions encompassed by a superconductor \cite{Sochnikov10}. Assuming fluxoid quantization, the observed oscillation period implies a characteristic geometrical area $A = (h/2e) (\Delta H)^{-1} = 0.643 \times 10^4~{\rm{nm}}^2$ and $1.44 \times 10^4~{\rm{nm}}^2$ for channels B1 and B2, respectively. This is far smaller than any relevant geometrical regions in our devices. The scaling of the oscillation period with the length of the \BS~channel (B1 and B2), roughly doubling when the channel length is halved, further rules out any insidious fluxoid quantization that might be associated with FIB damage in the contact region. 

Having dismissed standard explanations for our observations, we resort to a new model that predicts MR oscillations when a thin, narrow superconducting strip is placed in a perpendicular magnetic field \cite{Pekker,Johansson05}. We are cautiously aware that this ``Weber blockade'' model was developed for a genuine superconductor, while we are assuming that it can also be applied to a normal metal supporting proximity-induced superconductivity. There is theoretical basis for the formation of vortex-like pair correlations in the N region of diffusive SNS junctions \cite{Cuevas_2007}, but we are unaware of a similar prediction for a ballistic case such as the proximity effect in a high mobility TI surface state. In the Weber blockade model, the perpendicular magnetic field controls the number of Pearl vortices in the superconducting strip. Maximum dissipation (i.e. maximum resistance) occurs when the energies of configurations with $N$ and $N+1$ vortices are degenerate, and the MR then has a period related to the sequential addition of vortices with increasing magnetic field. Certain observations are immediately consistent with this model: for instance, the onset of periodic oscillations occurs only after a finite field (see insets to Figs. 6a and 6c) and the oscillation period is larger than that expected using fluxoid quantization in the geometric area of the channel. Assuming that a proximity induced superconductor can be described by a uniform superconducting coherence length $\xi$ (distinct from the length scale of the diffusion of Cooper pairs into the superconductor), this model can be extended to predict $\xi$ in the region of proximity induced superconductivity for our geometry. 

Using the geometrical parameters for our nanoribbons, we calculated the period of the MR oscillation as a function of the coherence length $\xi$ (i.e. the size of a vortex). Figure 6e shows the result of this calculation for channels B1 and B2, with channel width of 430 nm and nominal channel lengths of 0.94 \mum (squares) and 1.55 \mum (circles), respectively. A comparison with the experimentally observed period yields $\xi= 191 (186)$ nm for channel B1 (B2). This length scale corresponds to the size of a vortex in the \BS~channel and, as expected, is much larger than the coherence length of a typical superconductor (such as the tungsten contacts). The inset to Fig. 6e shows, in a model calculation, how vortices distribute in a nanoribbon as the energy (magnetic field) increases. In Fig. 6f, we plot the number of vortices in the channel as a function of the applied magnetic field, assuming the fitted values of $\xi$ in Fig. 6e. The model assumes a uniform superconducting gap; this may account for the disagreement between the experiment and the calculation using the fitted value of $\xi$ for the first few vortices (Fig. 6f).

\section{conclusions}
In summary, we have demonstrated proximity-induced superconductivity in \BS~nanoribbons contacted with superconducting electrodes. Our experiments suggest that even when diffusive bulk conduction coexists with ballistic surface conduction in a 3D TI, the proximity effect may preferentially couple to the latter. Furthermore, the superconductor-TI device configuration demonstrated here provides a viable route for creating vortices near the interface between these two classes of materials. The formation of these vortices is manifest in MR oscillations whose period can be well explained using a simple model. Thus, our experiments show that superconductivity -- and possibly vortices -- can be realized in experimental geometries that are directly relevant to the search for Majorana fermions in condensed matter systems \cite{Fu2008,Sau2010,Franz2011}.

\section{acknowledgement}
This work was supported by the Penn State Center for Nanoscale Science under the MRSEC program (NSF grant DMR-0820404). We also acknowledge partial support from ONR. We thank Liang Fu and David Pekker for insightful discussions.

\newpage
\begin{table}[ht]
\caption{Summary of \BS~nanoribbon devices}
\centering
\begin{tabular}{c c c c c c c} 
\hline\hline 
Device & Length & Width & Thickness & Contact width & Resistivity at 6 K & Contact material \\ [-1.5ex]
 & ($\mu$m) & (nm) & (nm) & (nm) & (m$\Omega\cdot$cm) &   \\ [0.5ex] 
\hline 
A & 1.08 & 600 & 60 & 200 & 0.51 & W \\
B1 & 0.94 & 430 & 60 & 430 & 1.37 & W \\
B2 & 1.55 & 430 & 60 & 430 & 1.37 & W \\
C & 2.47 & 284 & 276 & 316 & 0.79 & W \\
D & 2.30 & 240 & 77 & 240 & 0.70 & W \\
E & 5.12 & 300 & 98 & 122 & 0.69 & W \\
F & 1.00 & 700 & 60 & 300 & 3.53 & W \\
G & 5.73 & 270 & 60 & 234 & 0.39 & Pt \\
H & 0.94 & 370 & 50 & 280 & 0.48 & Pt \\ [1ex] 
\hline 
\end{tabular}
\label{Table1}
\end{table}

\newpage
\begin{figure}
\includegraphics[width=4in]{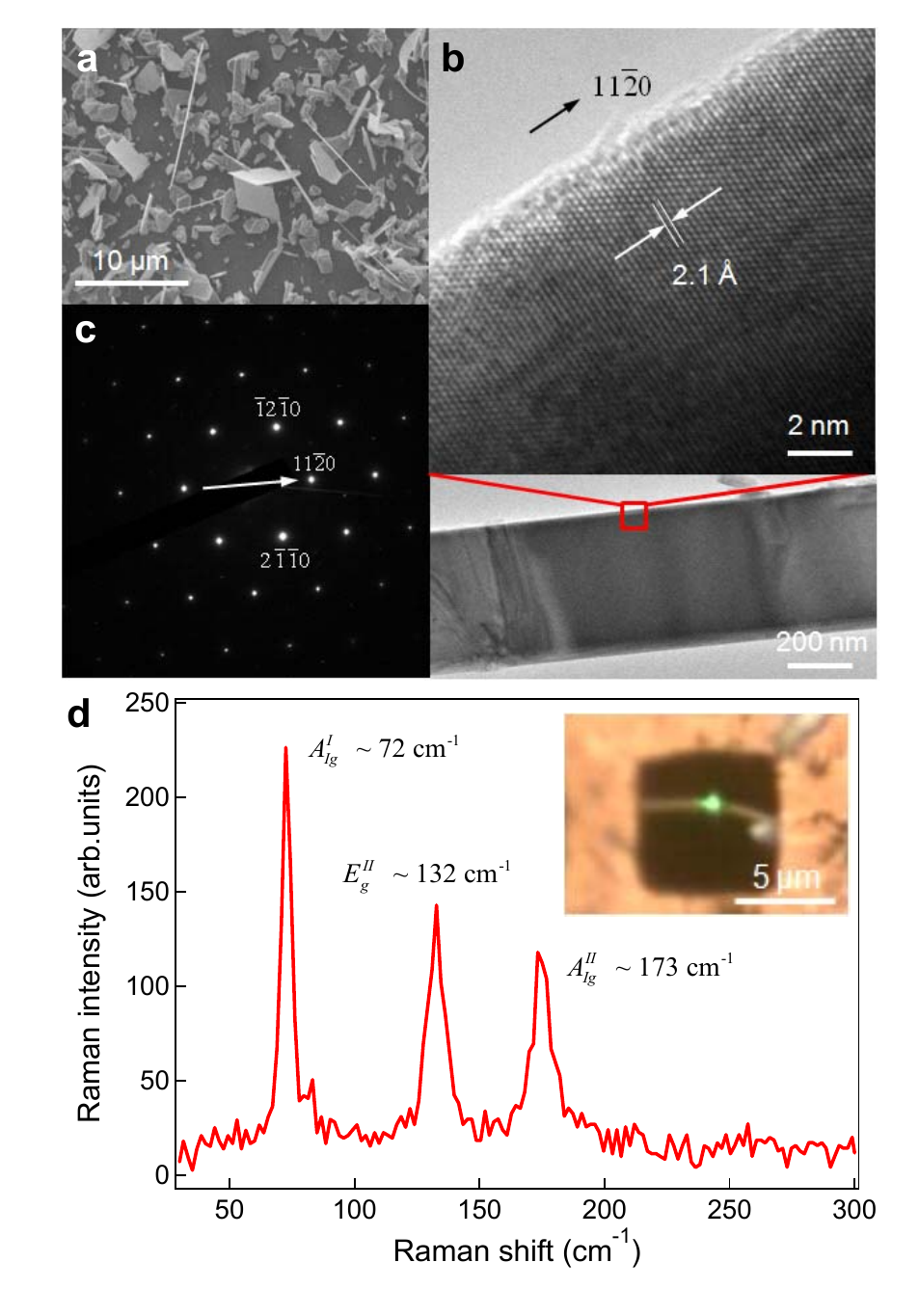} 
\caption{(Color online) {\bf a.} SEM image of as-grown \BS~ nanoribbons on a Si substrate. {\bf b.} HRTEM image from the edge of a \BS~ nanoribbon.  The interplanar distance along the growth direction $[11\bar{2}0]$ is 0.21 nm. {\bf c.} Selected area electron diffraction pattern from the nanoribbon in Fig. 1b shows  hexagonal symmetry.  The growth direction is along $[11\bar{2}0]$. {\bf d.} A typical Raman  spectrum from a single \BS~ nanoribbon with 514.5 nm excitation at room  temperature. All three vibration modes are consistent with those from  bulk \BS. The inset shows an optical image of a nanoribbon supported over a hole of a TEM grid. The green spot is the laser illumination.}
\label{Fig1}
\end{figure}

\newpage
\begin{figure}
\includegraphics[width=4in]{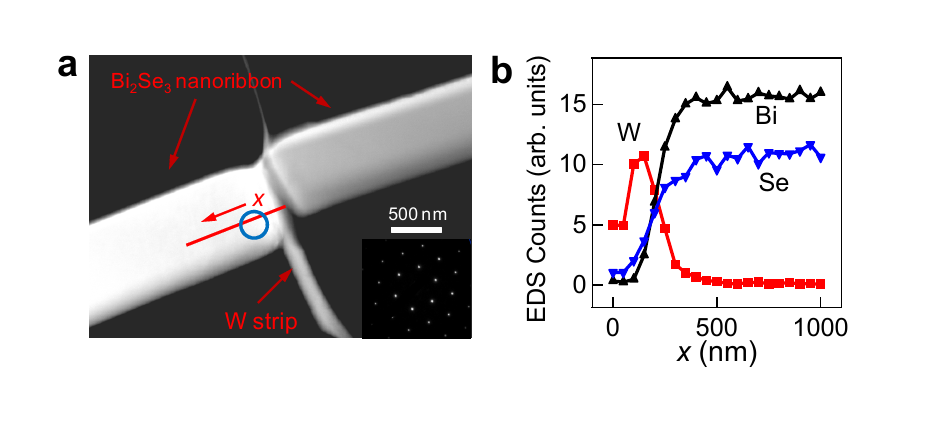} 
\caption{(Color online) {\bf a.} High-angle annular dark field image of a W strip deposited on a \BS~nanoribbon supported on a holey carbon TEM grid using the same deposition conditions for the actual devices. In the region next to the W contact (indicated by the circle), selected area diffraction pattern (inset) shows clear hexagonal \BS~single crystal structure. The line on the nanoribbon indicates the position of the X-ray energy dispersive line profile and the arrow shows the line scan direction. EDS spectra separated by 50 nm were collected in STEM mode using drift correction. {\bf b.} EDS counts (arb. units) vs. position plot of elemental W, Bi, and Se. The W signal is within the background at $\sim 450$ nm, which corresponds to a spread of $\sim 250$ nm from the left visible edge of the tungsten contact.}
\label{Fig. 2}
\end{figure}

\newpage
\begin{figure}
\includegraphics[width=4in]{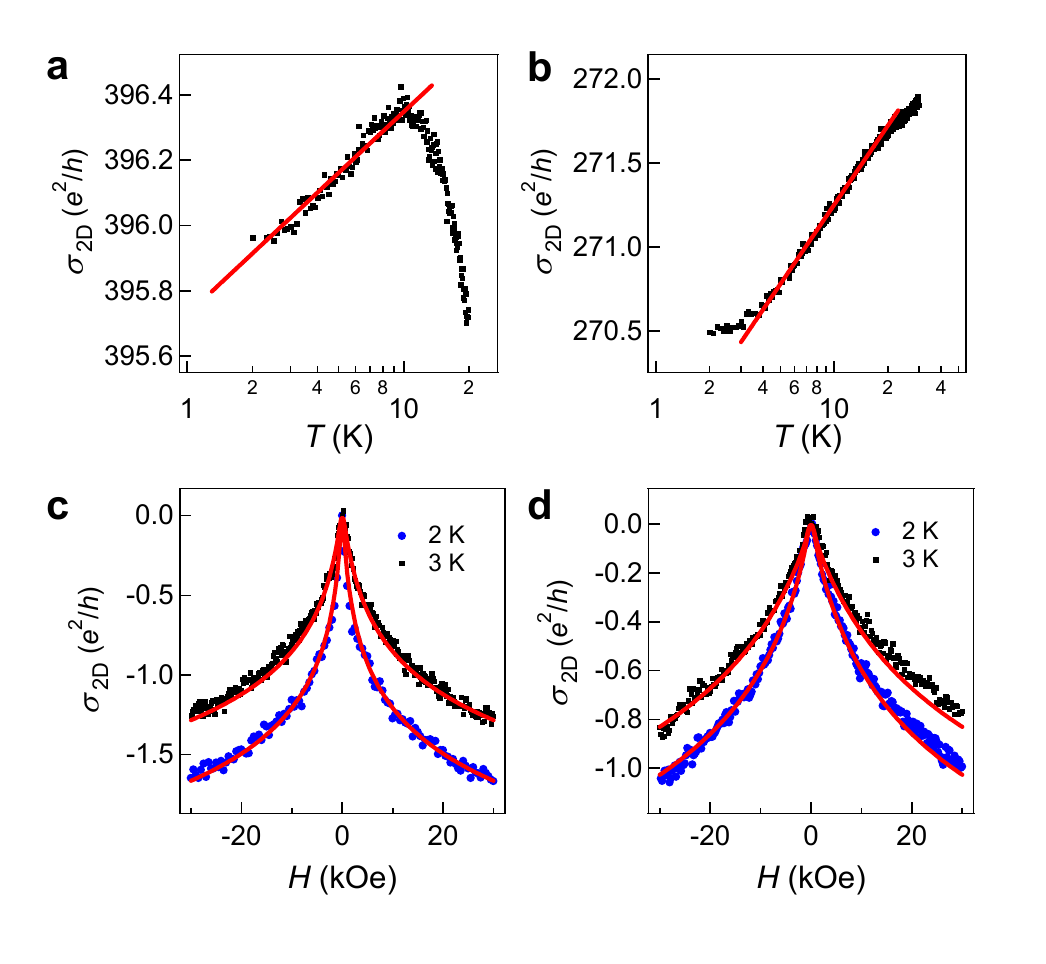}
\caption{(Color online) {\bf a.} Temperature dependent conductivity for device G. The solid line is a $\ln (T)$ fit. {\bf b.}Temperature dependent conductivity for device H. The solid line is a $\ln (T)$ fit. {\bf c.}Magnetoconductivity for device G at $T = 3.0$ K (squares) and $T = 2.0$ K (circles). The solid lines are fits to the Hikami-Larkin-Nagaoka theory with electron-electron interaction included. {\bf d.}Magnetoconductivity for device H at $T = 3.0$ K (squares) and $T = 2.0$ K (circles).  The solid lines are fits to the Hikami-Larkin-Nagaoka theory with electron-electron interaction.}
\label{Fig. 3}
\end{figure}

\newpage
\begin{figure}
\includegraphics[width=4in]{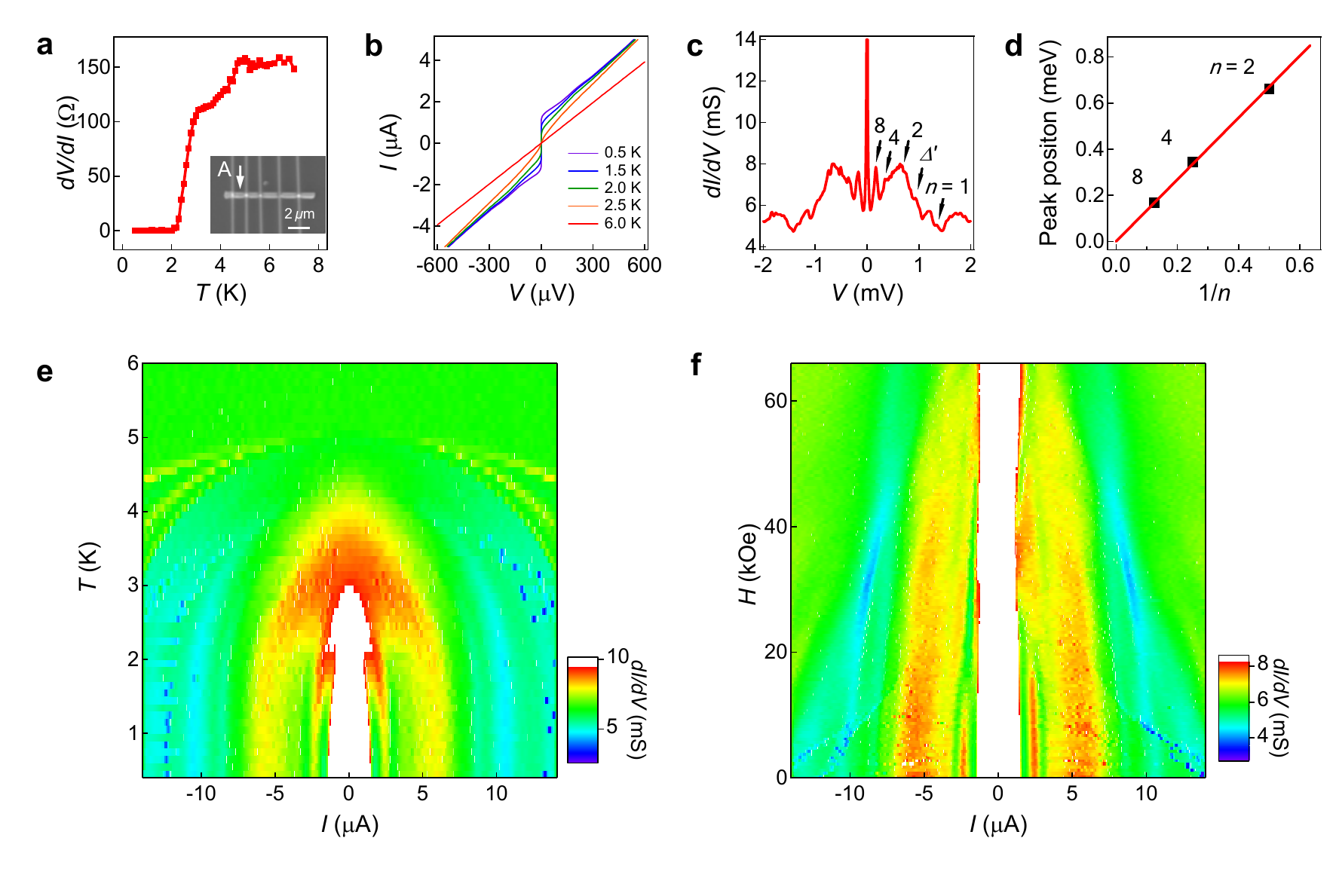}
\caption{(Color online) {\bf a.} Two probe zero bias $dV/dI$ vs. temperature for device A at $H = 0$. Inset shows an SEM image of the device, with the arrow indicating the measured channel with edge-to-edge length of 1.08 $\mu$m between two W electrodes. {\bf b.} $I-V$ characteristics of device A at various temperatures, measured using the same contacts as in (a). {\bf c.} $dI/dV$ vs. $V$ in device A at $T = 500$ mK and in zero magnetic field. The arrows identify a consistent subharmonic series of conductance anomalies corresponding to subharmonic gap structure ($\frac{2\Delta}{ne}$ with $n = 2, 4, 8$). {\bf d} Position of differential conductance anomalies as a function of the index $1/n$. {\bf e} Surface plot of $dI/dV$ as a function of the current $I$ and temperature at $H = 0$.  {\bf f} Surface plot of $dI/dV$ as a function of the current $I$ and perpendicular magnetic field $H$ at $T = 500$ mK. }
\label{Fig. 4}
\end{figure}

\newpage
\begin{figure}
\includegraphics[width=4in]{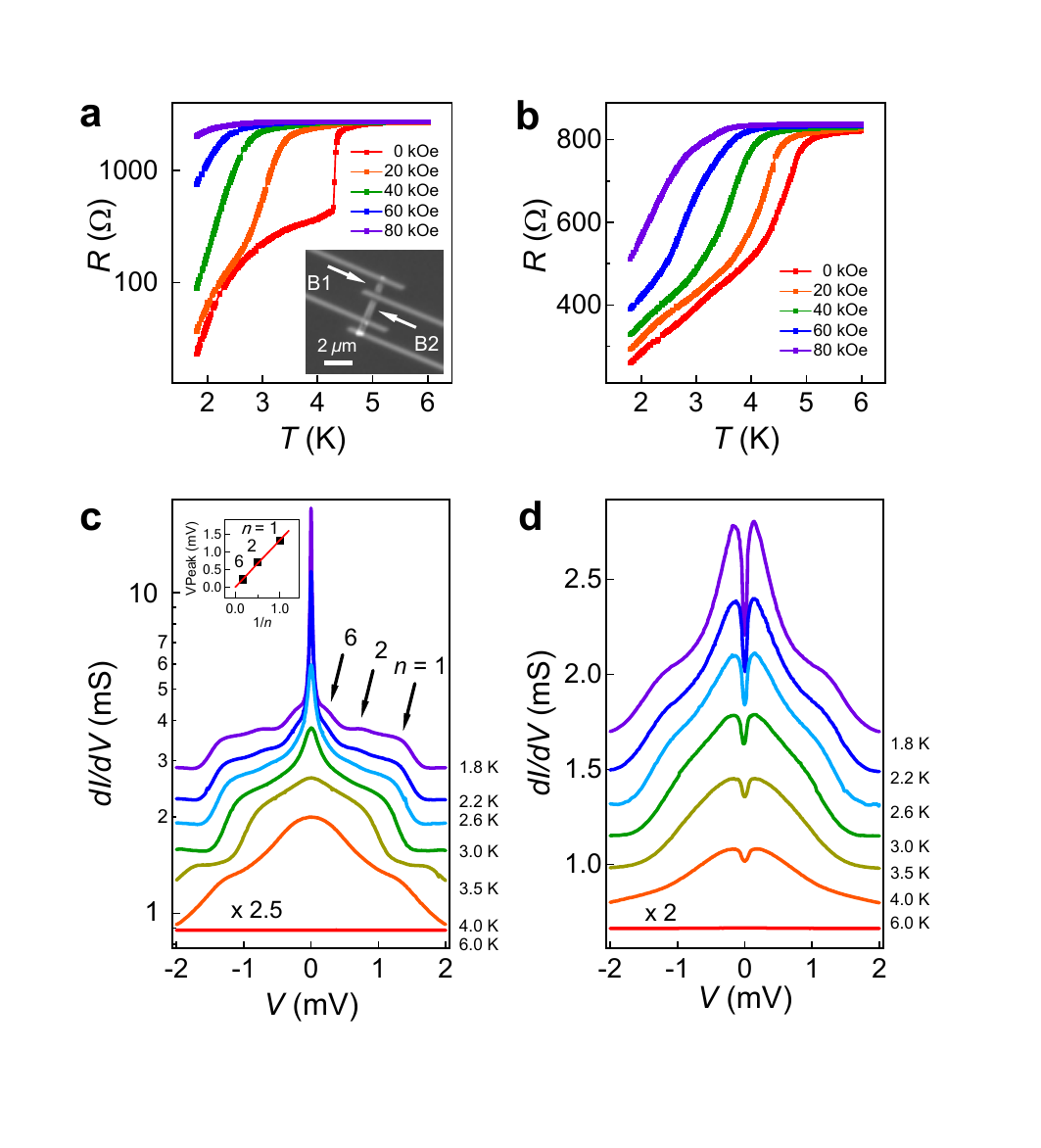}
\caption{(Color online) {\bf a.} Two probe resistance vs. temperature for channel B1 in different perpendicular fields. Inset shows an SEM image of the device B, with the two arrows indicating channels B1 and B2 with edge-to-edge length of 0.94 $\mu$m and 1.55 $\mu$m, respectively, between two W electrodes. {\bf b.} Four probe resistance vs. temperature for channel B2 in different perpendicular fields. {\bf c.} $dI/dV$ vs. $V$ for channel B1 at different temperatures and in zero magnetic field using a two probe configuration. Data for different temperatures are shifted or scaled with respected to the curve at $T = 4.0$ K for clarity. The arrows indicate conductance peaks corresponding to subharmonic gap structure ($2\Delta/ne$ with $n = 1, 2, 6$). {\bf d.} $dI/dV$ vs. $V$ for channel B2 at different temperatures and in zero magnetic field using a two probe configuration. The data for different temperatures are shifted or scaled with respected to the curve at $T = 4.0$ K for clarity.}
\label{Fig. 5}
\end{figure}

\newpage
\begin{figure}
\includegraphics[width=4in]{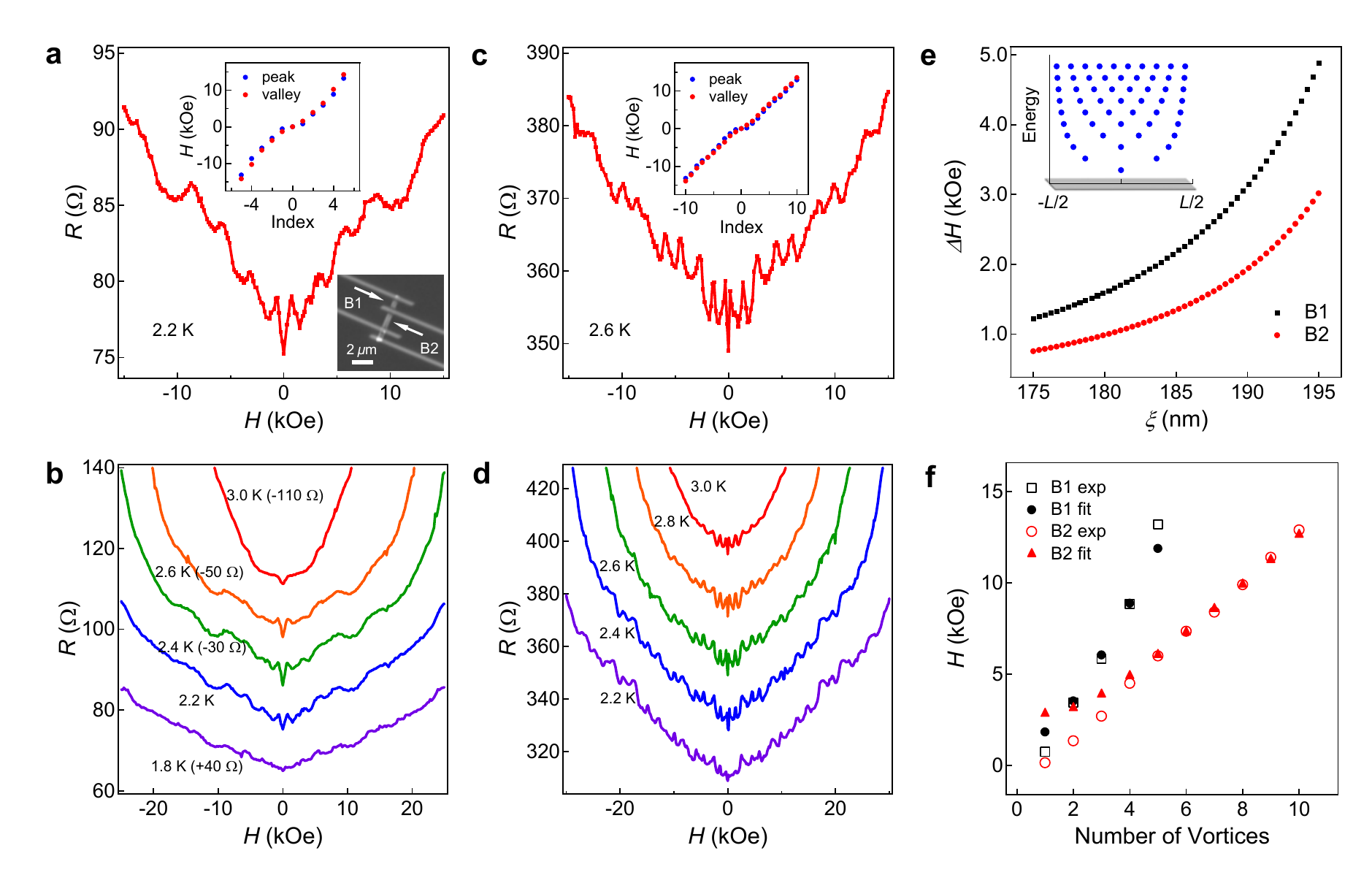}
\caption{(Color online) {\bf a.} MR in device B (channel B1, see lower inset) at $T = 2.2$ K. The upper inset plots the position of both peaks and valleys. {\bf b.} MR in channel B1 at different temperatures. The data are shifted with respect to the data at $T = 2.2$ K for clarity. {\bf c.} MR in channel B2 at $T = 2.6$ K. The inset plots the position of both peaks and valleys. {\bf d.} MR in channel B2 at different temperatures. The magnetic field in panels {\bf a-d} is perpendicular to the nanoribbon plane. {\bf e.} Calculated MR oscillation period vs. coherence length (size of vortex) for channels B1 (squares) and B2 (circles). Comparison with the data in Figs. 2a and 2c yields $\xi = 191$ nm and $\xi = 186$ nm for channel B1 and B2, respectively. The inset is a model calculation showing how vortices distribute in a nanoribbon as the energy (magnetic field) increases. {\bf f.} Magnetic field vs. number of vortices. Open symbols are experimental data and filled symbols are fits using the values of $\xi$ obtained in panel {\bf e}.}
\label{Fig. 6}
\end{figure}

\newpage
\begin{figure}
\includegraphics[width=4in]{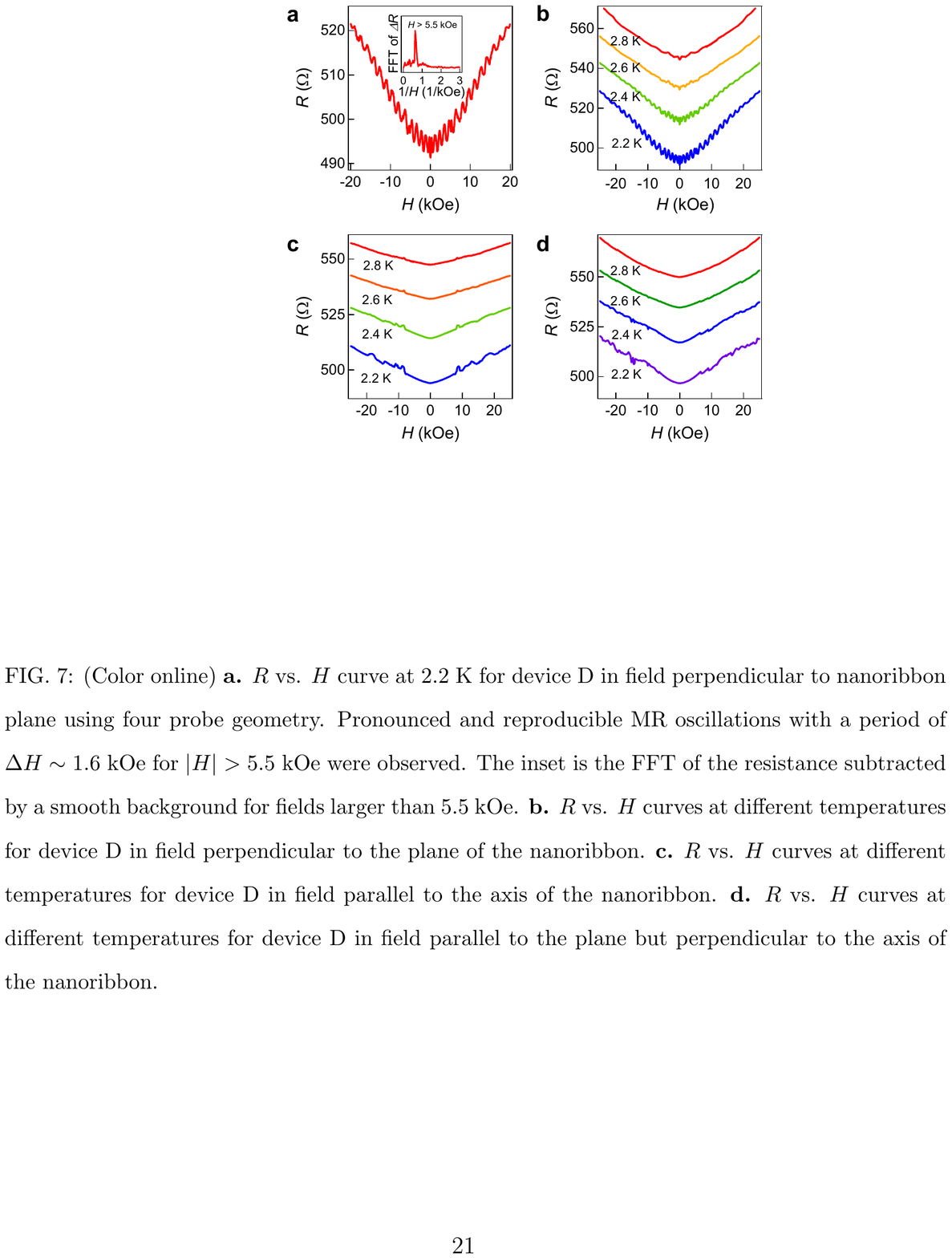}
\caption{(Color online) {\bf a.} MR for device D at $T=2.2$ K in field perpendicular to nanoribbon plane using four probe geometry. Pronounced and reproducible MR oscillations with a period of $\Delta H \sim 1.6$ kOe for $|H| > 5.5$ kOe were observed. The inset is the FFT of the MR after a smooth background is subtracted for $H > 5.5$ kOe. {\bf b.} MR at different temperatures for device D in field perpendicular to the plane of the nanoribbon. {\bf c.} MR at different temperatures for device D in field parallel to the axis of the nanoribbon. {\bf d.} MR at different temperatures for device D in field parallel to the plane but perpendicular to the axis of the nanoribbon.}
\label{Fig. 7}
\end{figure}

\end{document}